\documentclass[12pt,letterpaper]{article} 
\usepackage{amsthm,amsfonts,amsmath} 
\usepackage[pdftex]{graphicx}
\usepackage{graphicx}
\usepackage[square,sort]{natbib} 
\usepackage{epstopdf} 
\usepackage{appendix} 

\renewcommand{\labelenumi}{(\roman{enumi})} 
 
\newcommand{\be}{\begin{equation}} 
\newcommand{\ee}{\end{equation}} 
\newcommand{\bew}{\begin{equation*}} 
\newcommand{\eew}{\end{equation*}} 
 
\newcommand{\hhi}{{\rm HHI}} 
\newcommand{\rdi}{{\rm RDI}} 
\newcommand{\rc}{{\rm RC}}

\begin{document} 

\renewcommand{\labelenumi}{(\arabic{enumi})} 

\title{\large \textbf{Risk Without Return }\normalsize} 
\author{Lisa R. Goldberg\footnote{ Department 
of Statistics, University of California, 
Berkeley, CA 94720-3880, USA, email: {\tt lrg@stat.berkeley.edu}.} 
\and Ola Mahmoud\footnote{Pictet Wealth Management, 60 Route des Acacias, CH-1211 Geneva 73, Switzerland, and Coleman Fung Risk Management Research Center, University of California, Berkeley, CA 94720-3880, USA, email: {\tt omahmoud@pictet.com}.}} \normalsize 
\date{ \today\footnote{This research was supported by the Coleman Fung Risk Management Research Center at University of California, Berkeley.  We are grateful to Bob Anderson, Stephen Bianchi and Michael Hayes for their contributions to this article.}\\ \bigskip
 In {\it Journal of Investment Strategies}}
\maketitle 

\begin{abstract} 

Risk-only investment strategies have been growing in popularity as traditional investment strategies have fallen short of return targets over the last decade. However, risk-based investors should be aware of four things. First, theoretical considerations and empirical studies show that apparently distinct risk-based investment strategies are manifestations of a single effect. Second, turnover and associated transaction costs can be a substantial drag on return. Third, capital diversification benefits may be reduced. Fourth, there is an apparent connection between performance and risk diversification. To analyze risk diversification benefits in a consistent way, we introduce the Risk Diversification index ($\rdi$)  which measures risk concentrations and complements the Herfindahl-Hirschman index ($\hhi$) for capital concentrations.  
\\[2ex] 
\end{abstract} 

\renewcommand{\baselinestretch}{1.1} \small \normalsize 

In an inverted caricature of  pre-Markowitz investing, some funds are allocating assets strictly on the basis of risk and without regard for expected return. Strategies based on minimum variance, beta and risk parity have been growing, both in popularity and assets under management, as traditional investment approaches have fallen short of return targets over the last decade. Risk-only strategies are not a new idea. \citet{portfolio} identifies the minimum variance portfolio as optimal for a mean-variance investor whose estimates of asset expected returns are all equal.

\paragraph{Risk With Return} 

The Capital Asset Pricing Model (CAPM) predicts a linear relationship between the expected excess return of a portfolio and its market beta. 
However, empirical studies show that this simple relationship is not correct. In a seminal article, \citet{bjs} document the first CAPM anomaly: risk-adjusted returns of high-beta equities are too low and risk-adjusted returns of low-beta equities are too high by the standards of the CAPM. 
Two decades later, \citet{fama-french-92} find that size and value factors add to the explanation of stock return provided by market beta. 
The Fama--French three-factor model is, by far, the most well-established CAPM alternative. 
As in the case of the low-beta anomaly, there is disagreement about the underlying drivers of size and value effects, and there are contributions to the literature from both behavioral and neoclassical finance. 
In a survey of the vast empirical literature on the CAPM, \citet{ffcapm} comment: 
\begin{quote} 
The conflict between the behavioral irrational pricing story and the rational risk story for the empirical failures of the CAPM leave us at a timeworn impasse. 
\end{quote} 
However, there may be a connection between risk-based investing and the expected-return-based Fama--French model. \citet{scherer} regresses returns to a minimum variance portfolio onto the size and value factors. He finds that ``83\% of the variation of the minimum variance portfolio excess returns (relative to a CAPM alternative) can be attributed to the Fama--French factors":
\begin{quote} 
...investors can achieve a higher Sharpe ratio than the minimum variance portfolio by directly identifying the risk based pricing anomalies that the minimum variance portfolio draws upon
\end{quote} 
and he makes provocative comment:
\begin{quote} In this author's view, the minimization of risk is ---on its own---a meaningless objective. 
\end{quote} 
To what extent do recent data support this remark? Through a study of three popular risk-only strategies, we will show that risk-only can be a meaningful investment approach outperforming equally-weighted and balanced portfolios in terms of return and risk diversification.

\paragraph{Risk-Only Strategies}

Over the horizon from January 1988 to December 2010, we evaluate three popular risk-only strategies:  minimum variance, risk parity and low beta,  
based on four asset classes: US Equity, US Treasury Bonds, US Investment Grade Corporate Bonds, and Commodities \footnote{All four asset class time series are obtained from the Global Financial Data database (www.globalfinancialdata.com). We took the Russell 3000 Total Return Index, the USA 10-Year Government Bond Total Return Index, the USA Total Return AAA Corporate Bond Index, and the Goldman Sachs Commodity Price Index to represent US Equities, US Treasuries, US Investment Grade Corporate Bonds, and Commodities, respectively. }. 
In the minimum variance strategy, an asset's weight  is proportional to the sum of its conditional dependencies with other assets: 
\begin{align}\label{formula_minvar} 
\omega_i \propto \sum_j \sigma^{-1}_{ij}, 
\end{align} 
where $\sigma^{-1}_{ij}$ denote the elements of the inverse of the covariance matrix.

In the risk parity strategy, assets are weighted so their ex post risk contributions  are equal. An asset's weight in the low beta strategy is  inversely proportional to its benchmark beta.   All strategies are fully invested and long only.  Asset weights in the strategies depend on variance and covariance estimates, which are calculated  
using a 36-month rolling window of trailing returns. Varying the estimation methodology by changing the length of the rolling window or the weighting scheme applied to the returns within this window did not alter our results substantially. All strategies are rebalanced monthly. 

We  also consider an equally weighted portfolio and our benchmark: a  balanced portfolio of our four asset classes. 
With the traditional 60/40 balanced allocation in mind, we chose 60/20/10/10 weights for equities, commodities, corporate bonds and treasuries.

\begin{figure}[t] 
\includegraphics[scale = 0.93]{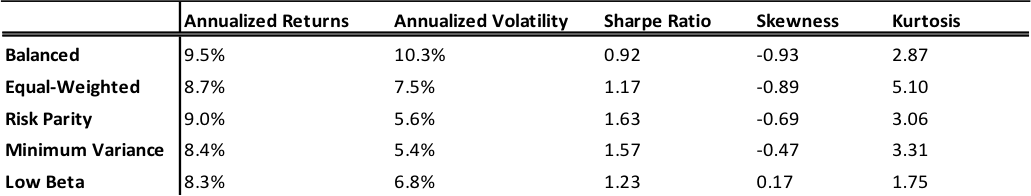} 
\caption{Summary statistics for the, balanced, equally-weighted, and risk-based strategies over the period January 1988--December 2010.} 
\label{strategy_stats} 
\end{figure}

Figure~\ref{strategy_stats} displays performance statistics on all strategies. Our benchmark has the highest annualized returns but also the highest volatility. From the perspective of risk and risk-adjusted returns, all three risk-only strategies beat the benchmark and the equal-weighted portfolio. 
Among the risk-based strategies, risk parity outperforms its rival minimum variance and low beta portfolios in terms of both return and Sharpe ratio.


\paragraph{Variations on a Theme} 


\begin{figure} 
\begin{center} 
\includegraphics[scale = 0.6]{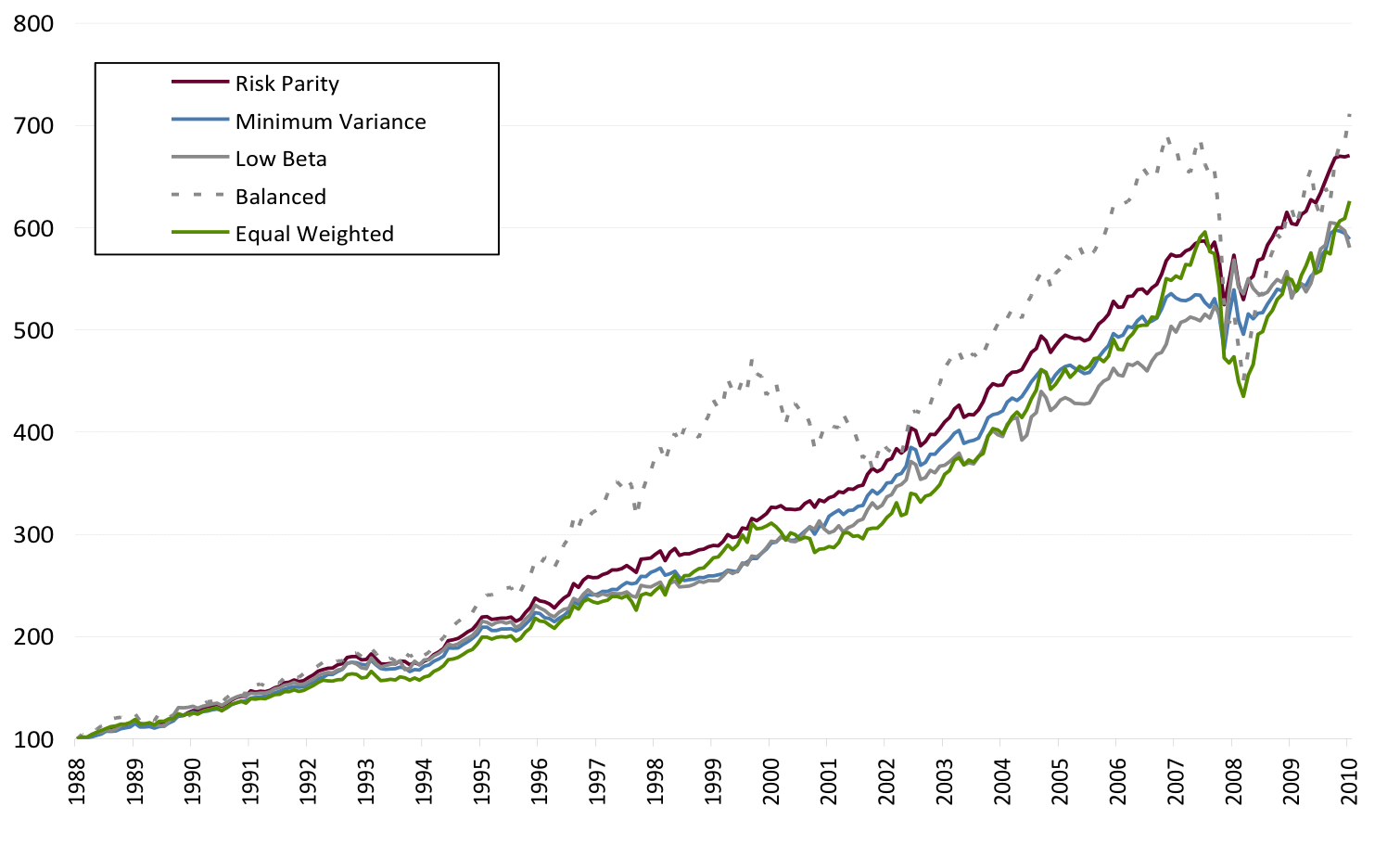} 
\end{center} 
\caption{Cumulative return to the balanced, equally-weighted and risk-based strategies over the period January 1988--December 2010.} 
\label{strategy_nav} 
\end{figure} 


\begin{figure} 
\begin{center} 
\includegraphics[scale = 1]{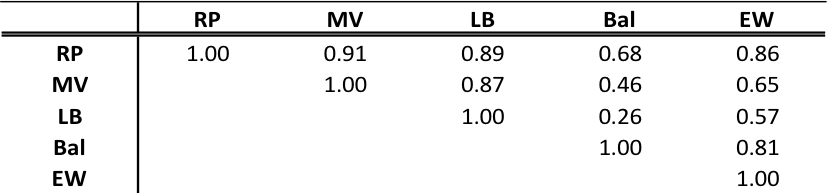} 
\end{center} 
\caption{Correlations between the balanced, equally-weighted and risk-based strategies over the period January 1988--December 2010.} 
\label{strategy_corr} 
\end{figure}

Some of the common elements to risk-based investment strategies are elucidated in \citet{minvar}, who provide a re-expression of Formula~(\ref{formula_minvar}) in a market with a single risk factor: 
\begin{align}\label{formula_minvarbeta} 
	\omega_n \propto \frac{1}{\sigma^2_{\epsilon n }}\left(1 - \frac{\beta_n}{\beta}\right). 
\end{align} 

Formula~\ref{formula_minvarbeta} shows that the weight of asset $n$ decreases as either its CAPM market model beta, $\beta_n$, or its idiosyncratic variance, $\sigma^2_{\epsilon i}$. increases. Further, if $\beta_n$ exceeds a threshold $\beta$, the weight of asset $n$ in the minimum variance portfolio is negative. Formula~(\ref{formula_minvarbeta}) suggests that portfolios emphasizing minimum variance, low-beta assets or assets with low volatility may be correlated. 

The performance of our three risk-only strategies over the horizon January 1988 to December 2010  provide empirical support for this hypothesis.  Figure~\ref{strategy_nav} shows cumulative returns to the three low-risk strategies and the balanced portfolio. We observe  co-movement across the low-risk strategies, with risk parity outperforming the others. Note that the benchmark has the highest return over the 22-year window; however it is also the most volatile strategy and it  has lowest Sharpe ratio. The correlations between strategies range from 0.87 to 0.95, as shown in Figure~\ref{strategy_corr}.

\paragraph{The Impact of Turnover} 


\begin{figure} 
\begin{center} 
\includegraphics[scale = 0.6]{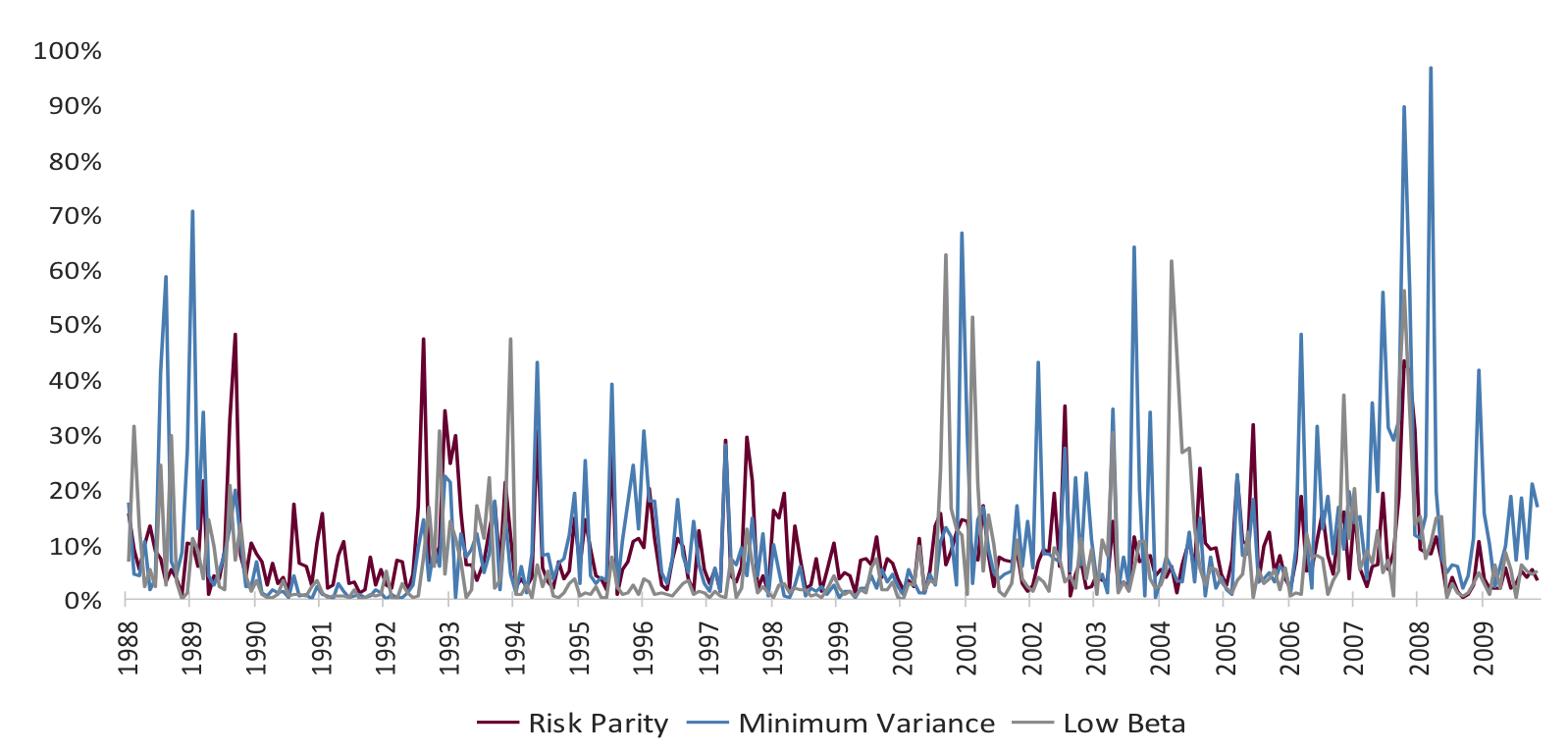} 
\end{center} 
\caption{Monthly turnover for the risk-based strategies over the period January 1988--December 2010.} 
\label{turnover} 
\end{figure}


\begin{figure} 
\begin{center} 
\includegraphics[scale = 0.6]{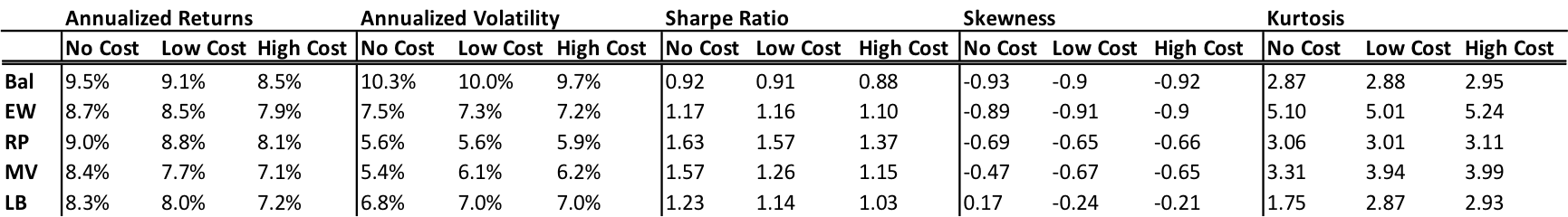} 
\end{center} 
\caption{Summary statistics for the balanced, equally-weighted and risk-based strategies with and without accounting for turnover-induced transaction costs over the period January 1988--December 2010.} 
\label{strats_with_turnover} 
\end{figure}


\begin{figure} 
\includegraphics[scale = 0.93]{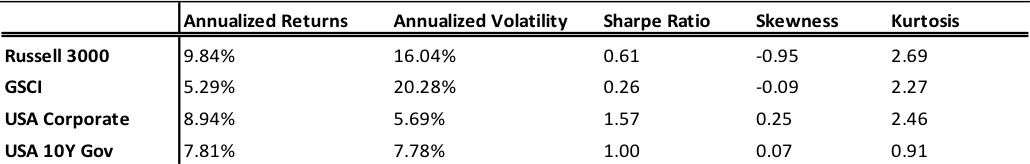} 
\caption{Summary statistics for the four asset classes used in the risk-based strategies over the period January 1988--December 2010.} 
\label{asset_stats} 
\end{figure}

We evaluate the drag on return caused by turnover-induced transaction costs.  Portfolio turnover from month to month is the lower of the total amount of assets sold and bought. Figure~\ref{turnover} shows this value over the 22-year period for each of the three risk-based portfolios. The average turnover is highest for minimum variance, followed closely by low beta, and then by risk parity. We observe that turnover is typically less than 10\% during bull markets, but often exceeds 50\% during turbulent market periods. This implies potential large rebalancing-induced trading costs in turbulent periods. 
Determining the precise relationship between turnover and its induced cost is beyond the scope of this paper,  so we simulate. We assume that turnover incurs a penalty in the form of a cost of 10 basis points on the lower end and 50 basis points on the higher end multiplied by the turnover. Summary statistics for the risk-based strategies and the benchmarks incorporating turnover-induced trading costs are shown in Figure~\ref{strats_with_turnover}. While the risk profile does not change substantially, cost diminishes returns and  Sharpe ratios, and the effect is more severe for the risk-only strategies.\footnote{The risk-only strategies that we consider are fully invested.  In practice, however, some risk-only strategies are levered.  \cite{abg} show that financing costs  can negate the outperformance of a levered risk parity strategy.}

\paragraph{Concentration Risk} 


\begin{figure} 
\begin{center} 
\includegraphics[scale = 0.4]{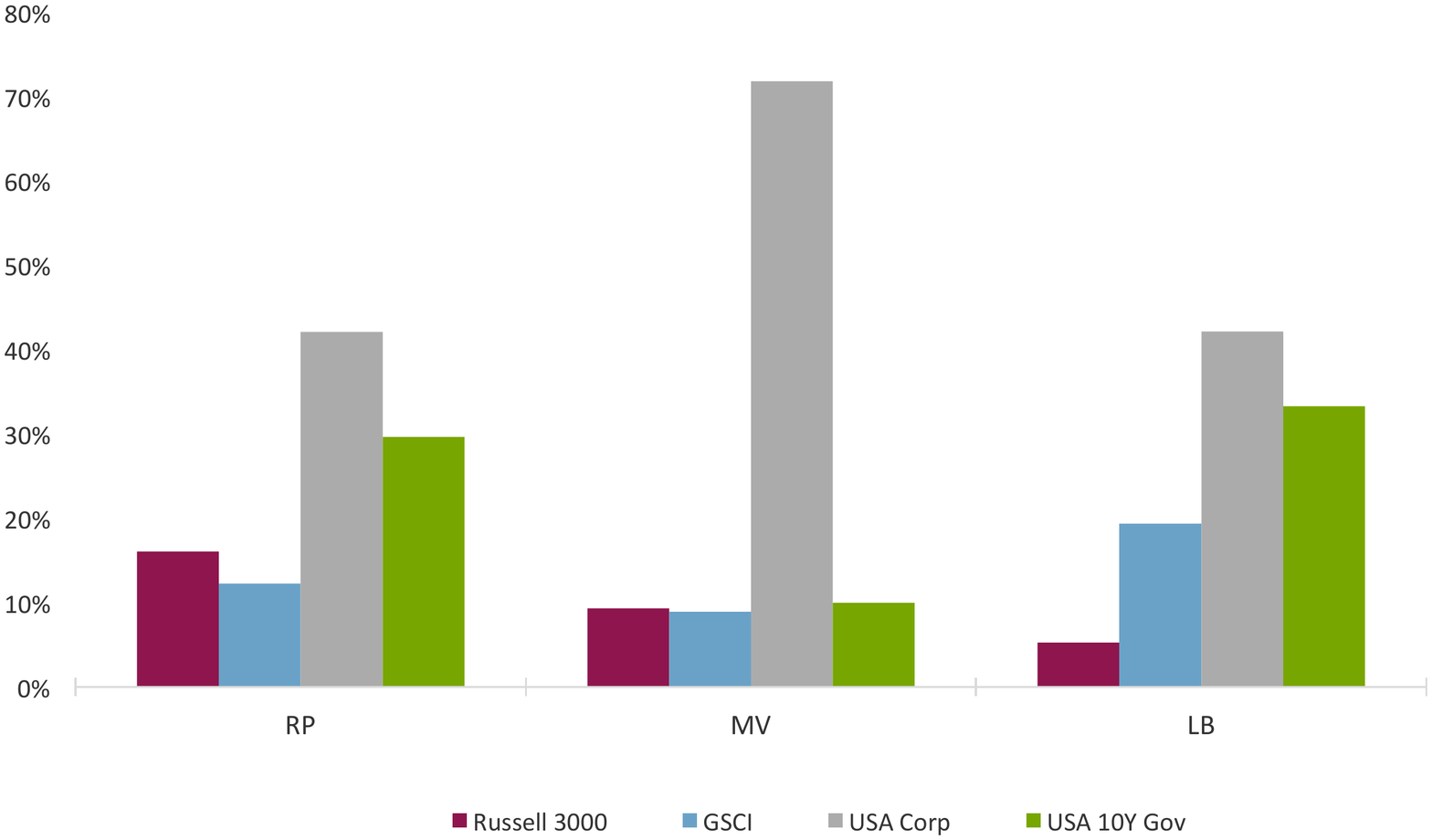} 
\end{center} 
\caption{Average capital allocations of risk-based strategies to the four asset classes over the period January 1988--December 2010.} 
\label{weights} 
\end{figure}

Low-risk strategies naturally concentrate on defensive assets.    The summary statistics in Figure~\ref{asset_stats} confirm  that  fixed income instruments have  lower volatilities
than do equities and commodities.  Figure~\ref{weights}  shows that
in all three strategies, the largest allocation is to  US Corporate Bonds, followed by US Treasuries; each strategy allocates at least 70\% of  its capital to fixed income.

A standard measure of concentration risk is the
normalized Herfindahl-Hirschman index:
\begin{align}\label{formula:HHI}
\hhi = \frac{\left(\sum_{n=1}^N w_n^2\right)   - 1/N}{1  - 1/N},
\end{align}
where $N$ is the number of assets in the portfolio and $w_n$ is the weight of asset $n$.
The values of HHI  range from 0 for an equally weighted portfolio to 1 for a portfolio composed of a single asset.

For the balanced portfolio,  $\hhi = 0.7$, due to the 60\% allocation to equities.  Among our risk-only strategies, minimum variance has the highest concentration risk, $\hhi  = 0.6$, followed by low beta, $\hhi = 0.3$.This relatively high concentration risk can be explained by the fixed income allocations. Given the weights of risk parity of Figure~\ref{weights}, it may seem  puzzling that this portfolio turns out to be almost perfectly diversified with an HHI of 0.08. Rather than lowering overall risk, risk parity equalizes the contributions of assets to overall portfolio risk.  In the example we considered, the act of equalizing risk contributions served to approximately equalize capital contributions.

Inspired by Herfindahl-Hirschman index, the  parallel between capital and risk allocations suggests a simple measure of risk diversification.  It is natural to measure risk diversification in terms of  fractional risk contributions, which are discussed in \citet{ghmm}. 
The fractional risk contribution of asset class $n$ to portfolio volatility  is given by:
 $$\rc_n = \left(\frac{1}{\sigma}\right) w_n\partial \sigma/\partial w_n,$$ where $\sigma$ is the risk of the portfolio.  Then $\sum_{n = 1}^N \rc_n = 1$ and: 
\begin{align}\label{formula:RDItrue}
\rdi = \frac{\left(\sum_{n=1}^N  \rc_n^2\right)   - 1/N}{1  - 1/N}
\end{align}   is the analog to $\hhi$.  In this article, we base strategies on the volatility of asset classes rather than their risk contributions.  This amounts to the tacit assumption that correlations between pairs of asset classes are zero.  Formula~\ref{formula:RDI} is a simplified $\rdi$ that takes account of that assumption:
\begin{align}\label{formula:RDI}
\rdi =\frac{\left(\sum_{n=1}^N  w_n^4 \sigma_n^4\right)/\left(\sum_{n=1}^N  w_n^2 \sigma_n^2\right)^2  - 1/N}{1  - 1/N}.
\end{align}

\begin{figure} 
\begin{center} 
\includegraphics[scale = 0.5]{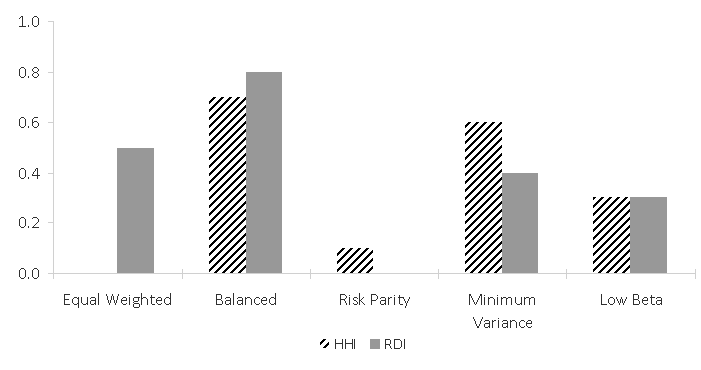} 
\end{center} 
\caption{Average concentration risk over the period January 1988--December 2010.} 
\label{figure:concentrations} 
\end{figure}

Figure~\ref{figure:concentrations} shows the Herfindahl-Hirschman and Risk Diversification indices for the strategies in our study. Only the risk parity strategy is diversified in both capital and risk.


\paragraph{Closing Remarks} 

Consistent with the large empirical literature on low-risk investing, we found that three risk-only strategies outperformed an equally weighted strategy  and a balanced strategy over the horizon from January 1988 to December 2010.  
There is no consensus about what drives the abnormal returns of  risk-based strategies exhibit in idealized settings, or whether these abnormal returns can reliably transcend the limitations to arbitrage. 
However, apparently distinct risk-based investment strategies are manifestations of a single effect, turnover and associated transaction costs can be a substantial drag on their return, and concentration is an understated source of  their risk. 

In our study,  risk parity outperformed the other strategies, and that may stem from the fact that the risk parity strategy was diversified both in capital and in risk weights.
Further research into the relationship between these two types of diversification is warranted.

\bibliographystyle{apalike} 
\bibliography{biblio} 

\end{document}